\documentclass{aa}  

\usepackage{graphicx}
\usepackage{txfonts}
\usepackage[colorlinks=true,citecolor=blue]{hyperref}
\usepackage{natbib}
\bibpunct{(}{)}{;}{a}{}{,} 
\begin{document}

   \title{
          Observed spectral energy distribution of the thermal emission from the dayside of \object{WASP-46b}
          \thanks{Based on observations collected with the Gamma Ray burst Optical 
                  and Near-infrared Detector (GROND) on the MPG/ESO 2.2~m telescope 
                  at the La Silla Observatory, Chile. Program 088.A-9016 (PI: Chen)}
          \fnmsep\thanks{Photometric time series are only available in electronic 
                         form at the CDS via anonymous ftp to cdsarc.u-strasbg.fr 
                         (130.79.128.5) or via http://cdsweb.u-strasbg.fr/cgi-bin/qcat?J/A+A/}}
   

   \author{G. Chen\inst{1,2,3}
          \and
           R. van Boekel\inst{3}
          \and
           H. Wang\inst{1}
          \and
           N. Nikolov\inst{3,4}
          \and
           U. Seemann\inst{5}
          \and
           Th. Henning\inst{3}
          }
   \institute{Purple Mountain Observatory \& Key Laboratory for Radio Astronomy, Chinese 
             Academy of Sciences, 2 West Beijing Road, Nanjing 210008, PR China\\
              \email{guochen@pmo.ac.cn}
         \and
             University of Chinese Academy of Sciences, 19A Yuquan Road, Beijing 100049, PR China
         \and
             Max Planck Institute for Astronomy, K\"onigstuhl 17, 69117 Heidelberg, Germany
         \and
             Astrophysics Group, University of Exeter, Stocker Road, EX4 4QL, Exeter, UK
         \and
             Institut f\"ur Astrophysik, Friedrich-Hund-Platz 1, 37077 G\"ottingen, Germany
             }

   \date{Received 12 March 2014; accepted 22 May 2014}
   \titlerunning{Secondary-eclipse observation of \object{WASP-46b}}
   \authorrunning{G. Chen et al.}

 
  \abstract
  {} 
   {We aim to construct a spectral energy distribution (SED) for the emission from the dayside atmosphere of the hot Jupiter \object{WASP-46b} and to investigate its energy budget.}
   {We observed a secondary eclipse of \object{WASP-46b} simultaneously in the $g'r'i'z'JHK$ bands using the GROND instrument on the MPG/ESO 2.2~m telescope. Eclipse depths of the acquired light curves were derived to infer the brightness temperatures at multibands that cover the SED peak. }
   {We report the first detection of the thermal emission from the dayside of \object{WASP-46b} in the $K$ band at 4.2$\sigma$ level and tentative detections in the $H$ (2.5$\sigma$) and $J$ (2.3$\sigma$) bands, with flux ratios of 0.253 $^{+0.063}_{-0.060}$\%, 0.194 $\pm$ 0.078\%, and 0.129 $\pm$ 0.055\%, respectively. The derived brightness temperatures (2306 $^{+177}_{-187}$~K, 2462 $^{+245}_{-302}$~K, and 2453 $^{+198}_{-258}$~K, respectively) are consistent with an isothermal temperature profile of 2386~K, which is significantly higher than the dayside-averaged equilibrium temperature, indicative of very poor heat redistribution efficiency. We also investigate the tentative detections in the $g'r'i'$ bands and the 3$\sigma$ upper limit in the $z'$ band, which might indicate the existence of reflective clouds if these tentative detections do not arise from systematics.}
   {}


   \maketitle

\section{Introduction}\label{sec:intro}

Observation of a hot-Jupiter secondary eclipse is one of the most efficient ways to characterize exoplanet atmospheres. The signals from the dayside atmosphere can be discerned through the comparison between in-eclipse and out-of-eclipse measurements. Such observations have been performed on dozens of exoplanets by the {\it Spitzer} and {\it Hubble} space telescopes in the mid-infrared and 1.1--1.7~$\mu$m wavelength range, respectively \citep[see the reviews by][and references therein]{2010ARA&A..48..631S,2014arXiv1402.1169M}. Chemical compositions, such as H$_2$O, CO, CO$_2$, and CH$_4$, can be investigated based on multiwavelength data with the help of an atmospheric retrieval modeling technique \citep[e.g.][]{2009ApJ...707...24M,2012MNRAS.420..170L,2012ApJ...749...93L}. Existence of thermal inversions is inferred using the emission behavior of these molecules \citep{2008ApJ...678.1419F,2008ApJ...678.1436B,2010ApJ...725..261M}. On the other hand, results from ground-based near-infrared observations have emerged rapidly during the past three years \citep[e.g.][]{2011AJ....141...30C,2011A&A...530A...5C,2011A&A...528A..49D,2012A&A...542A...4G,2012ApJ...748L...8Z,2012ApJ...744..122Z,2012ApJ...754..106D,2012ApJ...760..140C,2013A&A...550A..54D,2013ApJ...770...70W,2014ApJ...781..109O,2014A&A...564A...6C,2014A&A...563A..40C}, providing a good complement to the wavelength ranges covered by {\it Spitzer} and {\it Hubble}. The planetary energy budget can be constrained in the near-infrared where the planetary spectral energy distribution (SED) peaks. The vertical temperature profile can be constrained at the lower atmospheric layers since the $J$, $H$, $K$ bands overlap weaker molecular bands and thus probe deeper than the mid-infrared bands. 

\object{WASP-46b} is a hot Jupiter orbiting a G6V star every 1.43 days \citep{2012MNRAS.422.1988A}. Its mass and radius were deduced to be 2.10~$M_{\rm{Jup}}$ and 1.31~$R_{\rm{Jup}}$, resulting in a bulk density similar to our Jupiter. Taking the planetary radius, the orbital distance, and the stellar effective temperature into account, it is expected to be highly irradiated, with an incident flux of 1.7$\times$10$^9$ erg$^{-1}$\,cm$^{-2}$. This translates into an equilibrium temperature of 1654~K, assuming zero albedo and planet-wide heat redistribution. Combined with the high planet-to-star area ratio ($R_{\rm{p}}^2/R_{\star}^2=0.02155$), this hot Jupiter is a favorable target for detecting the thermal emission from its dayside atmosphere. \citet{2012MNRAS.422.1988A} also found that its host star is active, since it has a 16-day rotational modulation and weak \ion{Ca}{II} H+K emission. The stellar activity might prevent a thermal inversion in the upper atmosphere, given the hypothesis that the high UV flux could destroy the high-altitude compounds responsible for the formation of temperature inversion \citep{2010ApJ...720.1569K}. 

We aim at detecting the dayside emission of \object{WASP-46b} by observing its secondary eclipse simultaneously in seven bands. We expect to construct an SED for its dayside emission, to study its energy budget, and to obtain first insight into its temperature profile. This work belongs to our project that characterizes hot-Jupiter atmospheres using the technique of simultaneous multiband photometry, in which we have detected the thermal emission of \object{WASP-5b} in the $J$ and $K$ bands \citep{2014A&A...564A...6C} and the thermal emission of \object{WASP-43b} in the $K$ band \citep{2014A&A...563A..40C}. 

Here we report the first detection of the thermal emission from the dayside atmosphere of \object{WASP-46b}. This paper is organized as follows: in Sect.~\ref{sec:data} we summarize the observation and data reduction. In Sect.~\ref{sec:analysis} we describe the light-curve modeling process. In Sect.~\ref{sec:discuss} we present the analysis results and discuss the atmosphere delivered by our data. Finally, Sect.~\ref{sec:con} gives the conclusions.

\section{Observations and data reduction}\label{sec:data}

We observed one secondary eclipse of \object{WASP-46b} using the Gamma Ray burst Optical and Near-infrared Detector \citep[GROND;][]{2008PASP..120..405G} mounted on the MPG/ESO 2.2~m telescope at La Silla in Chile. This imaging instrument uses dichroics to acquire data simultaneously in the Sloan $g'r'i'z'$ and near-infrared (NIR) $JHK$ bands. The optical arm employs backside-illuminated $2048\times2048$ E2V CCDs without antiblooming structures, which has a field of view (FOV) of $5.4'\times5.4'$ (0$\farcs$158 per pixel). The NIR arm employs $1024\times1024$ Rockwell HAWAII-1 arrays, with an FOV of $10'\times10'$ (0$\farcs$60 per pixel). 

\begin{figure}
\centering
\includegraphics[width=0.9\hsize]{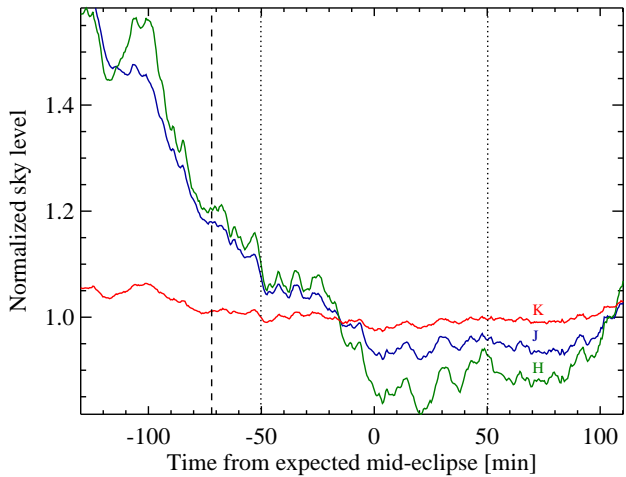}
\caption{Normalized sky level versus elapsed time. Blue, green, and red curves show the sky levels of the $J$, $H$, and $K$ bands. For each band, all the values are normalized by its median, so as to illustrate the overall variation. Data that were acquired before the time indicated by the vertical dashed line are not included in our final analysis. Two dotted lines indicate the expected first and last contacts of the secondary eclipse based on the assumption of a circular orbit.\label{fig:sky}}
\end{figure}

The secondary eclipse was observed on October 9 2011. The observation started at 90 minutes before the first contact (i.e. ingress) and ended at 50 minutes after the last contact (i.e. egress), continuously lasting four hours, with an overall airmass below 1.40. Staring mode was adopted and no defocusing was applied. We used 30~s exposures for the optical cameras and 4.5~s exposures for the NIR cameras. The optical images were collected in fast read-out mode. The NIR images were averaged every four frames before read-out. Most of the peak pixels of the target star in the $JHK$ bands have values around $10^4$~ADU, while a few can rise as high as $2\times10^4$~ADU at the end of the observation. All of these values lie below the 1\% nonlinearity of the detector, thus no nonlinearity correction was applied in the subsequent calibration to avoid introducing additional noise. After accounting for the dead time caused by telescope operation and the time for file read-out and saving, we calculated an effective duty cycle of 51\% and 61\% for the optical and NIR bands. Since this observation started before the end of evening twilight, the sky count levels were high at the beginning and decreased quickly afterwards. After checking the relative sky variation (see Fig.~\ref{fig:sky}), we discarded the first-hour data\footnote{The rejection criteria at $t=-0.05$~days is somewhat arbitrary. We tried to avoid including the data with higher sky levels. We also tried to keep part of pre-ingress data to constrain the light-curve decorrelation. The compromise was made at the adopted time criteria, where the slope of the sky values is steeper before this time.} in the subsequent reduction and analysis. This leaves 185 and 371 frames for the optical and NIR bands. 

The optical and NIR images were reduced following the approach described in \citet{2014A&A...564A...6C,2014A&A...563A..40C}. In brief, the calibration steps for the optical bands include bias subtraction and flat division; for the NIR bands, they are dark subtraction, readout pattern removal, and flat division. We performed aperture photometry on \object{WASP-46} and several reference stars using the IDL routines of DAOPHOT. The reference star ensembles for each band were selected to be unsaturated and to have similar brightness to \object{WASP-46}. They were individually normalized by each one's median flux and were then weighted combined as the composite reference. After dividing the flux of \object{WASP-46} by that of the composite reference, we constructed the eclipse light curves that were free of common-mode systematics to the first order. We empirically selected the optimal aperture radius and sky annuli sizes by minimizing the rms of the eclipse light curves. Consistent results were found for the photometry that is derived with aperture radius and sky annulus sizes close to the optimal choice. 

We extracted the UTC time stamps and centered them on the actual mid-exposure time. They were converted into BJD$_{\rm{TDB}}$ using the IDL routines written by \citet{2010PASP..122..935E}. Since the eclipse light curves show visible systematics (see the left-hand panels of Fig.~\ref{fig:occlc_opt} and \ref{fig:occlc_nir}) correlated with target location drift, seeing variation, and possibly airmass, we also collected the following auxiliary time-correlated information for the subsequent modeling process: target centroid on the detector ($x$, $y$), FWHM of the target's PSF ($s_x$, $s_y$), and airmass ($z$).

\section{Light-curve analysis}\label{sec:analysis}

   \begin{table}
     \small
     \centering
     \caption{Derived and adopted parameters of \object{WASP-46}}
     \label{tab:sys}
     \begin{tabular}{lccc}
     \hline\hline\noalign{\smallskip}
     Parameter & Units & Value & Ref.\\\noalign{\smallskip}
     \hline\noalign{\smallskip}
     $T_{\rm{mid,occ}}$    & BJD$_{\rm{TDB}}$ & 2455843.6050 $^{+0.0021}_{-0.0020}$ & 1\tablefootmark{b}\\\noalign{\smallskip}
     $\phi_{\rm{mid,occ}}$\tablefootmark{a} & ... & 0.5047 $^{+0.0014}_{-0.0014}$ & 1\tablefootmark{d}\\\noalign{\smallskip}
     $T_{\rm{offset}}$\tablefootmark{a}     & min & 9.6 $^{+3.0}_{-2.9}$ & 1\tablefootmark{d}\\\noalign{\smallskip}
     $T_{58}$             & days   & 0.0639 $^{+0.0049}_{-0.0044}$    & 1\tablefootmark{c}\\\noalign{\smallskip}
     $e\cos\omega$\tablefootmark{a}        & ... & 0.0073 $^{+0.0022}_{-0.0022}$ & 1\tablefootmark{d}\\\noalign{\smallskip}
     $e\sin\omega$        & ... & $-$0.018 $^{+0.015}_{-0.015}$ & 1\tablefootmark{d}\\\noalign{\smallskip}
     $e$        & ... & 0.020 $^{+0.014}_{-0.010}$ & 1\tablefootmark{d}\\\noalign{\smallskip}
     $\omega$   & degree & $-$68 $^{+47}_{-11}$ & 1\tablefootmark{d}\\\noalign{\smallskip}
     \hline\noalign{\smallskip}
     $T_{\rm{mid,tran}}$    & BJD$_{\rm{TDB}}$ & 2455392.316279 & 2\\\noalign{\smallskip}
     Period                 & days   & 1.43037    & 2\\\noalign{\smallskip}
     Inclination            & degree & 82.63      & 2\\\noalign{\smallskip}
     $a/R_{\star}$          & ...    & 5.74       & 2\\\noalign{\smallskip}
     $R_{\rm{p}}/R_{\star}$ & ...    & 0.146799   & 2\\\noalign{\smallskip}
     $T_{14}$               & days   & 0.06973    & 2\\\noalign{\smallskip}
     Eccentricity           & ...    & 0. (adopted)    & 2\\\noalign{\smallskip}
     $\omega$               & degree & 0. (adopted)    & 2\\\noalign{\smallskip}
     $\log g_{\star}$       & cgs    & 4.493 $\pm$ 0.023 & 2\\\noalign{\smallskip}
     $T_{\rm{eff,\star}}$   & K      & 5620 $\pm$ 160   & 2\\\noalign{\smallskip}
     $[$Fe/H$]$             & dex    & $-$0.37 $\pm$ 0.13 & 2\\\noalign{\smallskip}
     $T_{\rm{eq,pl}}(A=0,f=1/4)$       & K      & 1654 $\pm$ 50   & 2\\\noalign{\smallskip}
     \hline
     \end{tabular}
     \tablefoot{
        \tablefoottext{a}{Light travel time ($\sim$24.3 s) in the system has been 
                          corrected \citep{2005ApJ...623L..45L}.}
        \tablefoottext{b}{Derived from the original joint analysis described in 
                              Sect.~\ref{sec:analysis}.}
        \tablefoottext{c}{Derived from the analysis with Gaussian priors on $T_{\rm{mid,occ}}$ 
                              and $F_{\rm p}/F_\star$ that is also described in Sect.~\ref{sec:analysis}.}
        \tablefoottext{d}{Derived from the posterior probability distributions of the two analyses 
                              (b) and (c), see Sect.~\ref{sec:ecc}.}
                }
     \tablebib{(1) This work; (2) \citet{2012MNRAS.422.1988A}.}
   \end{table}

\begin{figure*}
\centering
\includegraphics[width=0.32\hsize]{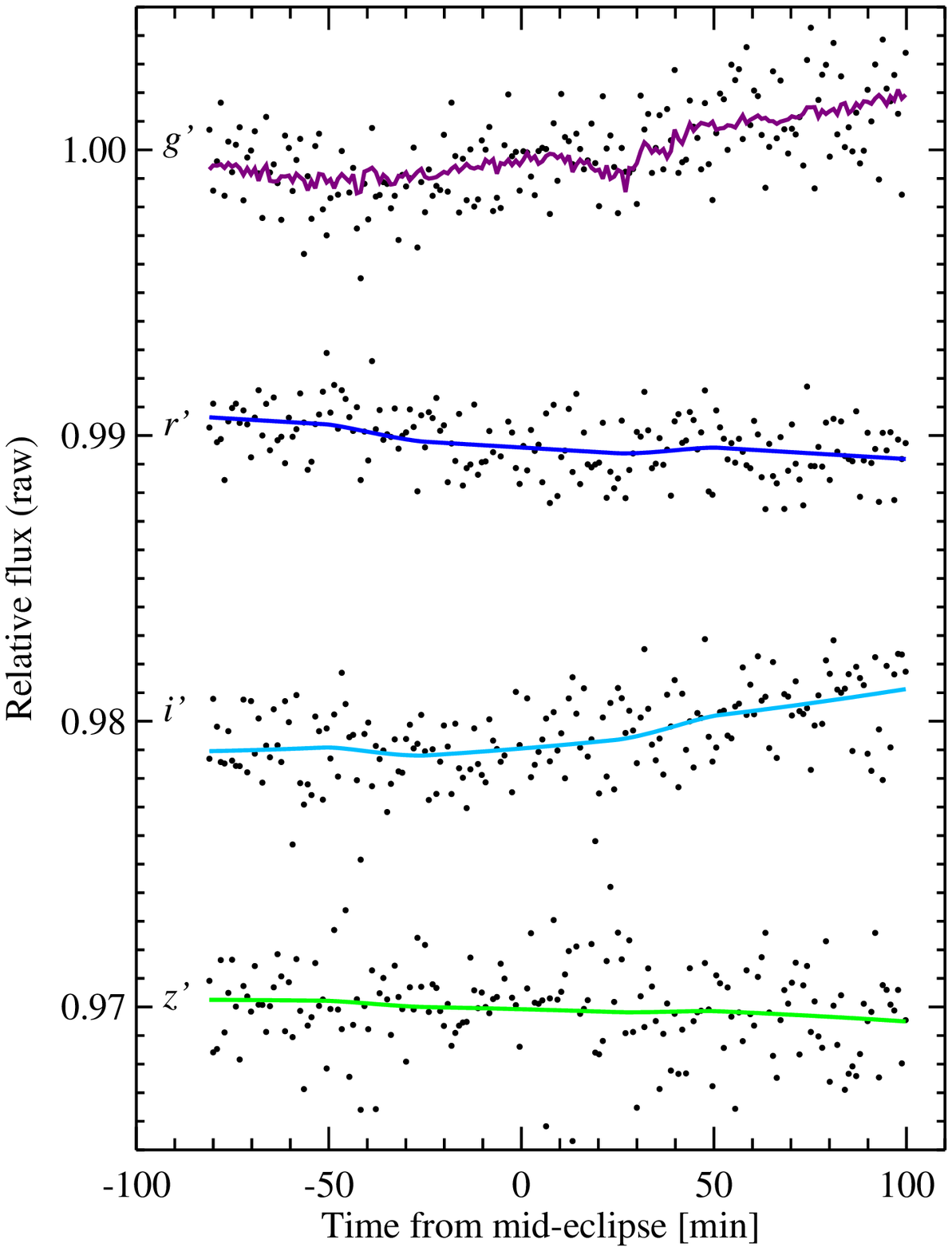}
\includegraphics[width=0.32\hsize]{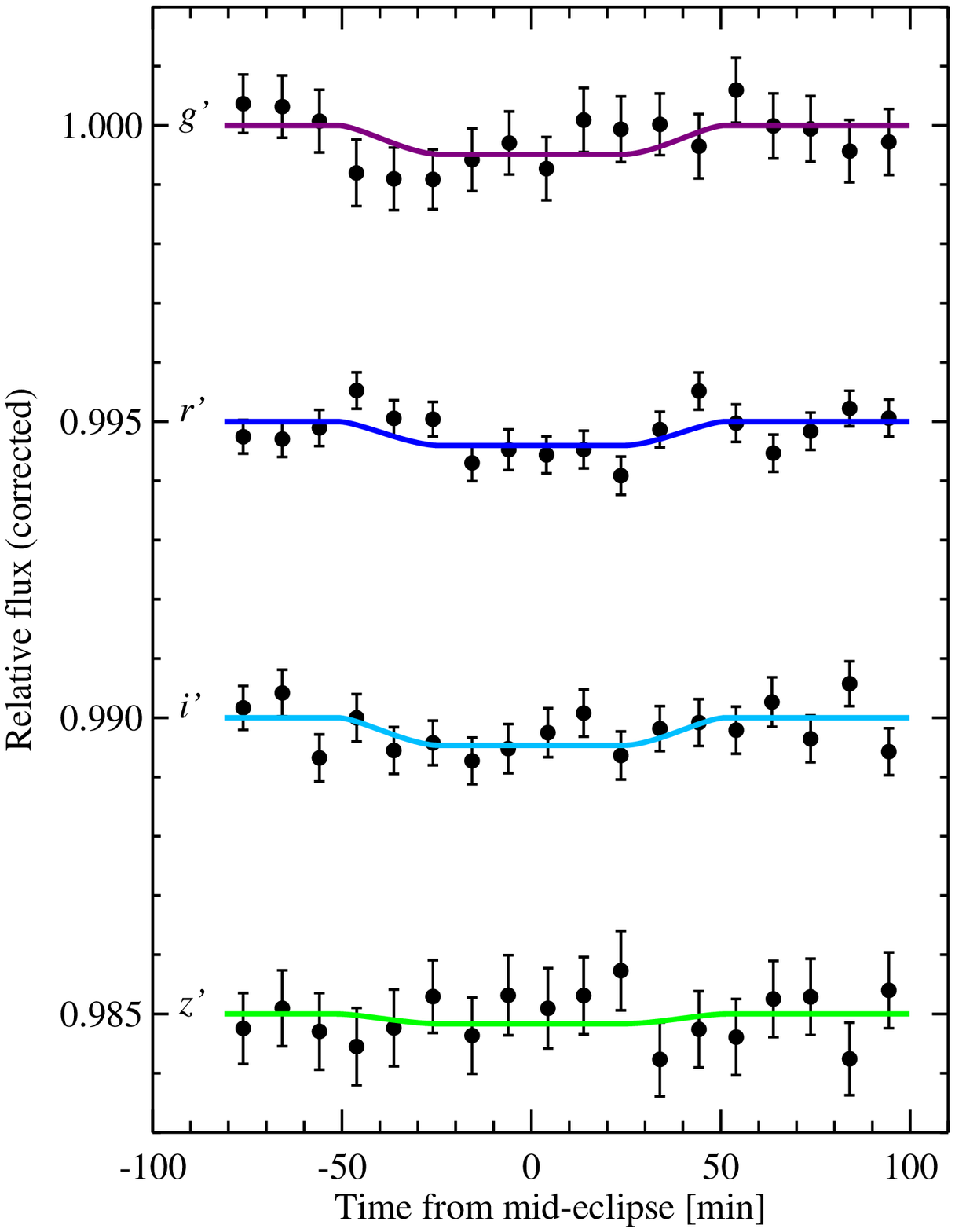}
\includegraphics[width=0.32\hsize]{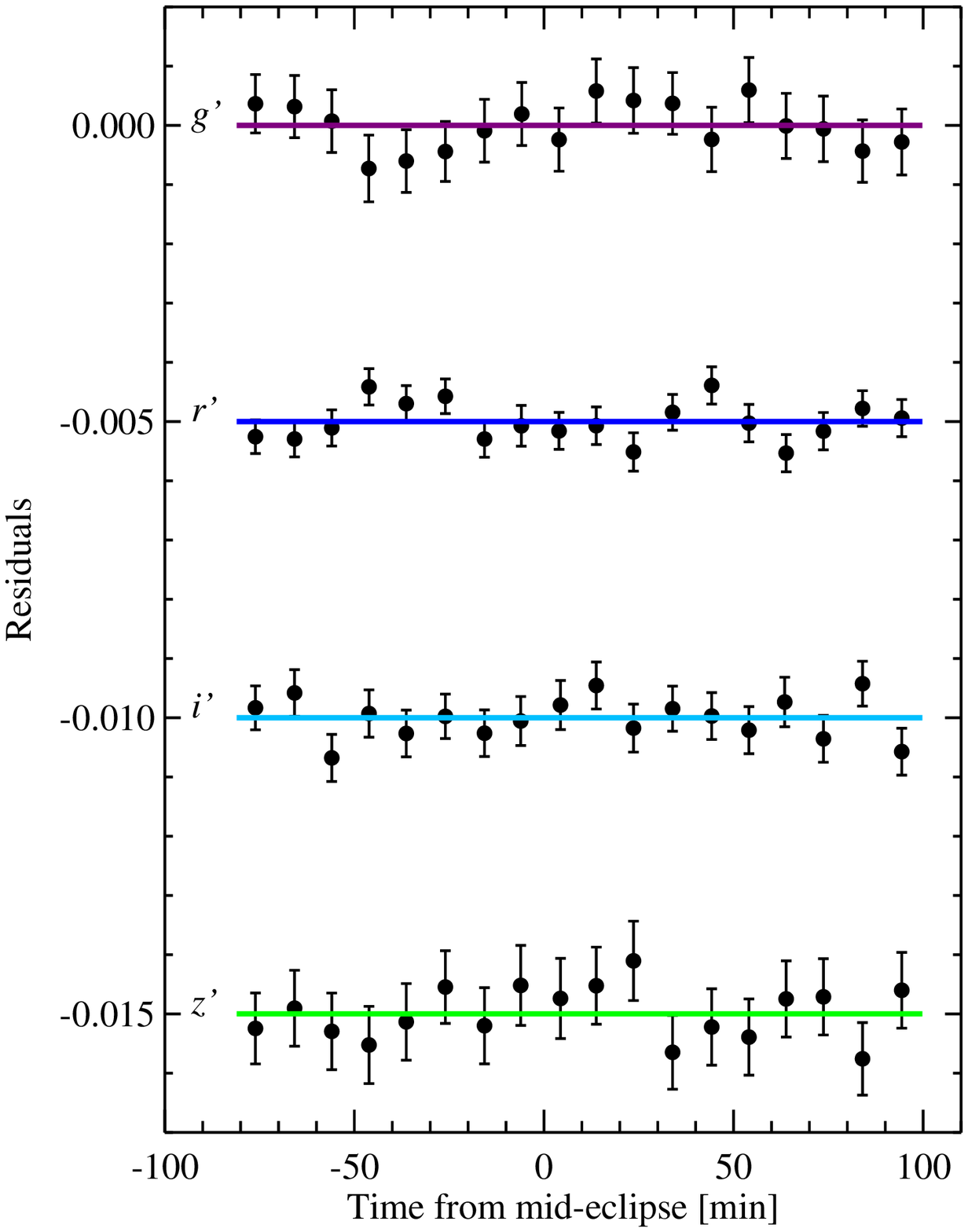}
\caption{Secondary-eclipse light curves of \object{WASP-46b} as observed with GROND in the $g'r'i'z'$ bands. {\it Left panel}: light curves that are normalized by the reference stars. {\it Middle panel}: baseline-corrected light curves that are binned every 10 minutes for display purpose. {\it Right panel}: best-fit light-curve residuals that are binned every 10 minutes. All the best-fits models are indicated by solid lines. \label{fig:occlc_opt}}
\end{figure*}

\begin{figure*}
\centering
\includegraphics[width=0.32\hsize]{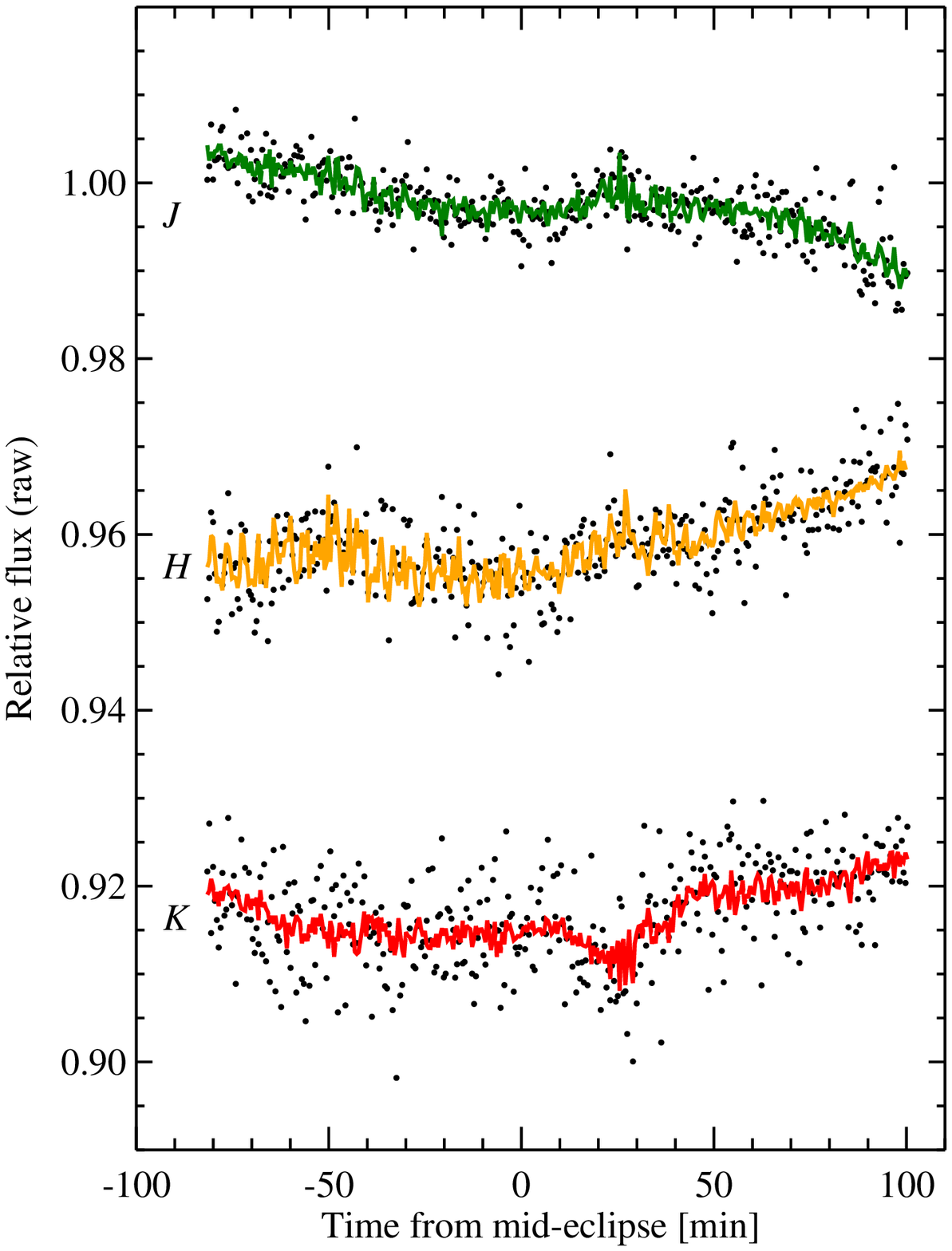}
\includegraphics[width=0.32\hsize]{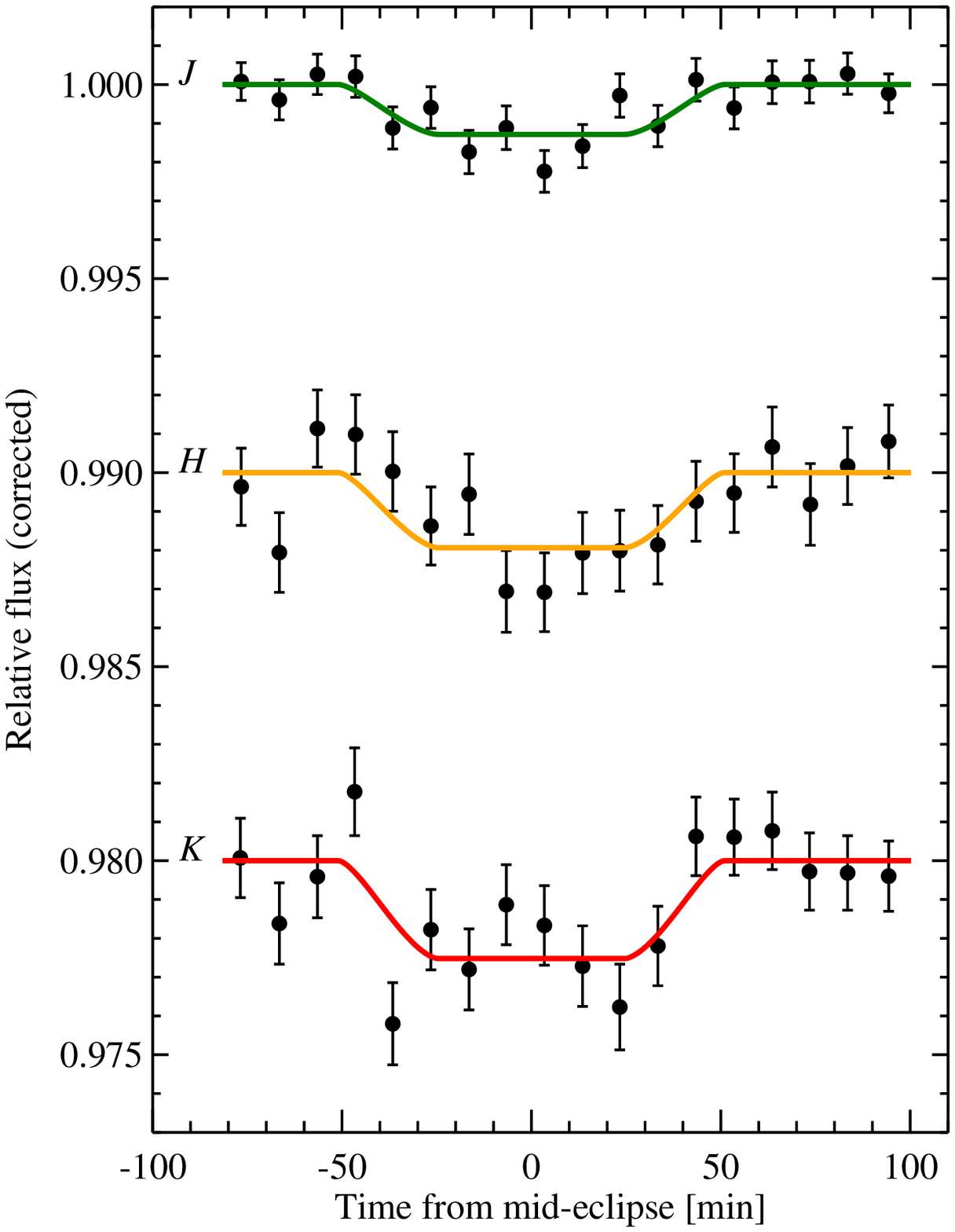}
\includegraphics[width=0.32\hsize]{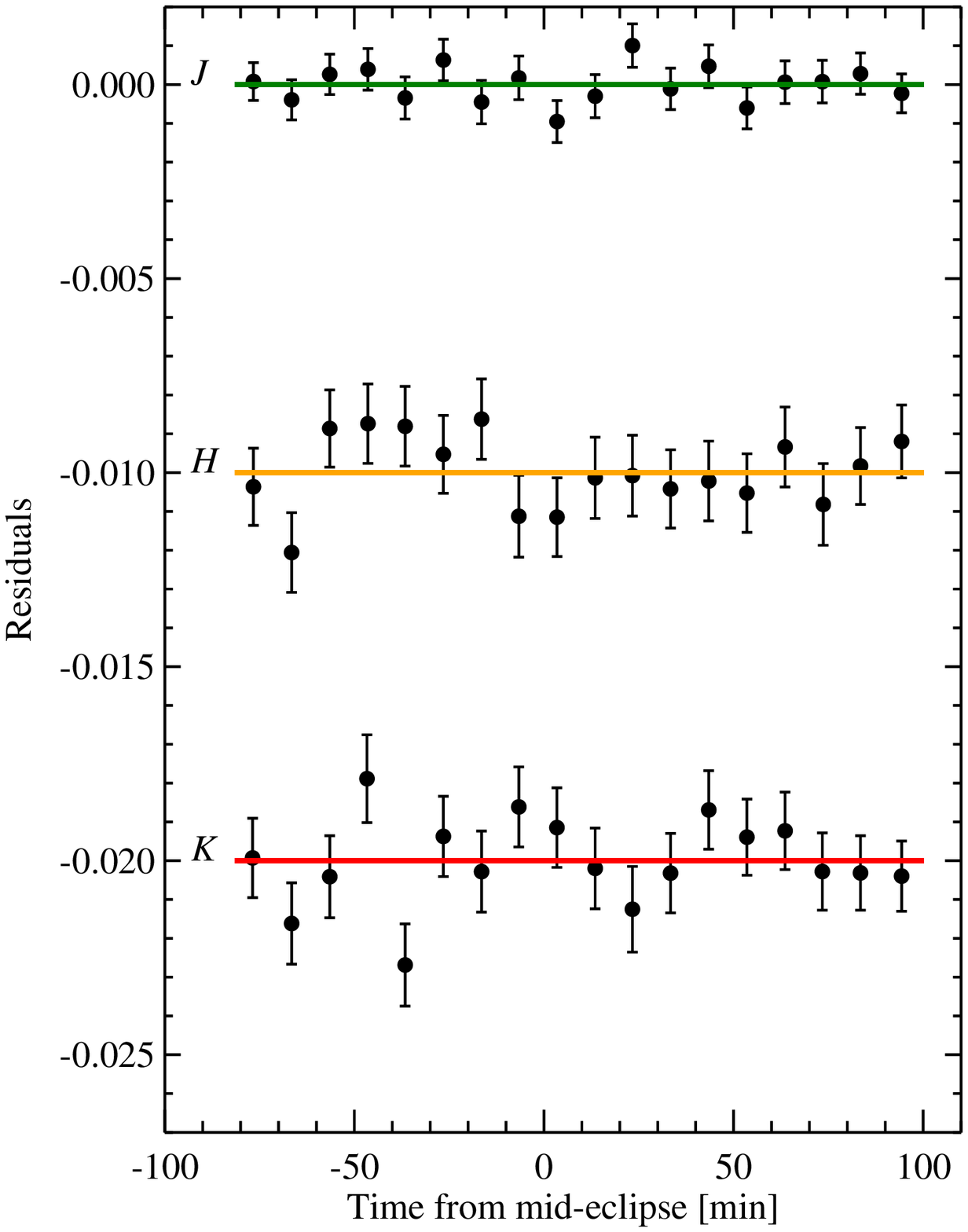}
\caption{Secondary-eclipse light curves of \object{WASP-46b} as observed with GROND in the $JHK$ bands. {\it Left panel}: light curves that are normalized by the reference stars. {\it Middle panel}: baseline-corrected light curves that are binned every 10 minutes for display purpose. {\it Right panel}: best-fit light-curve residuals that are binned every 10 minutes. All the best-fits models are indicated by solid lines. \label{fig:occlc_nir}}
\end{figure*}

\begin{figure*}
\centering
\includegraphics[width=0.28\hsize]{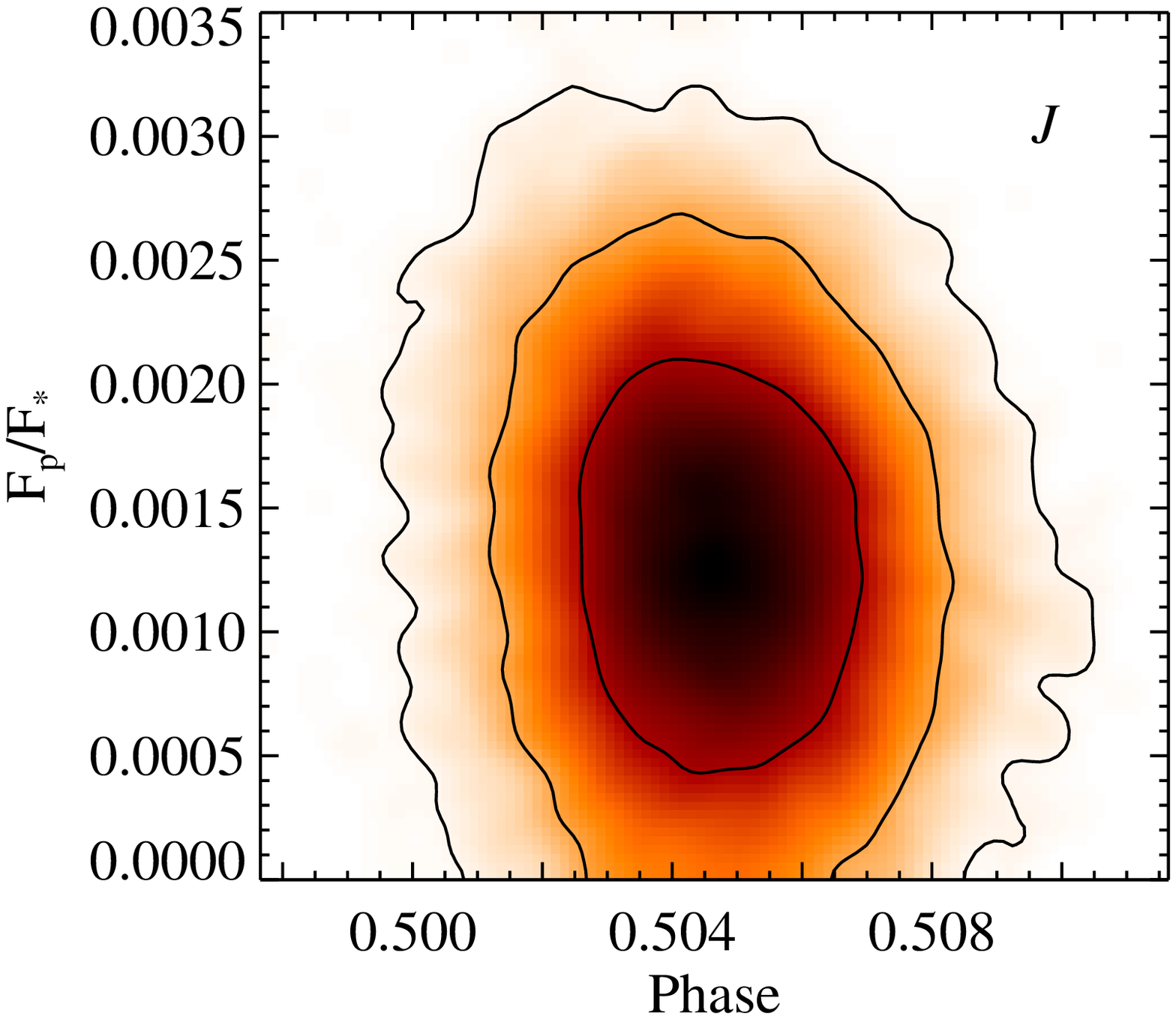}
\includegraphics[width=0.28\hsize]{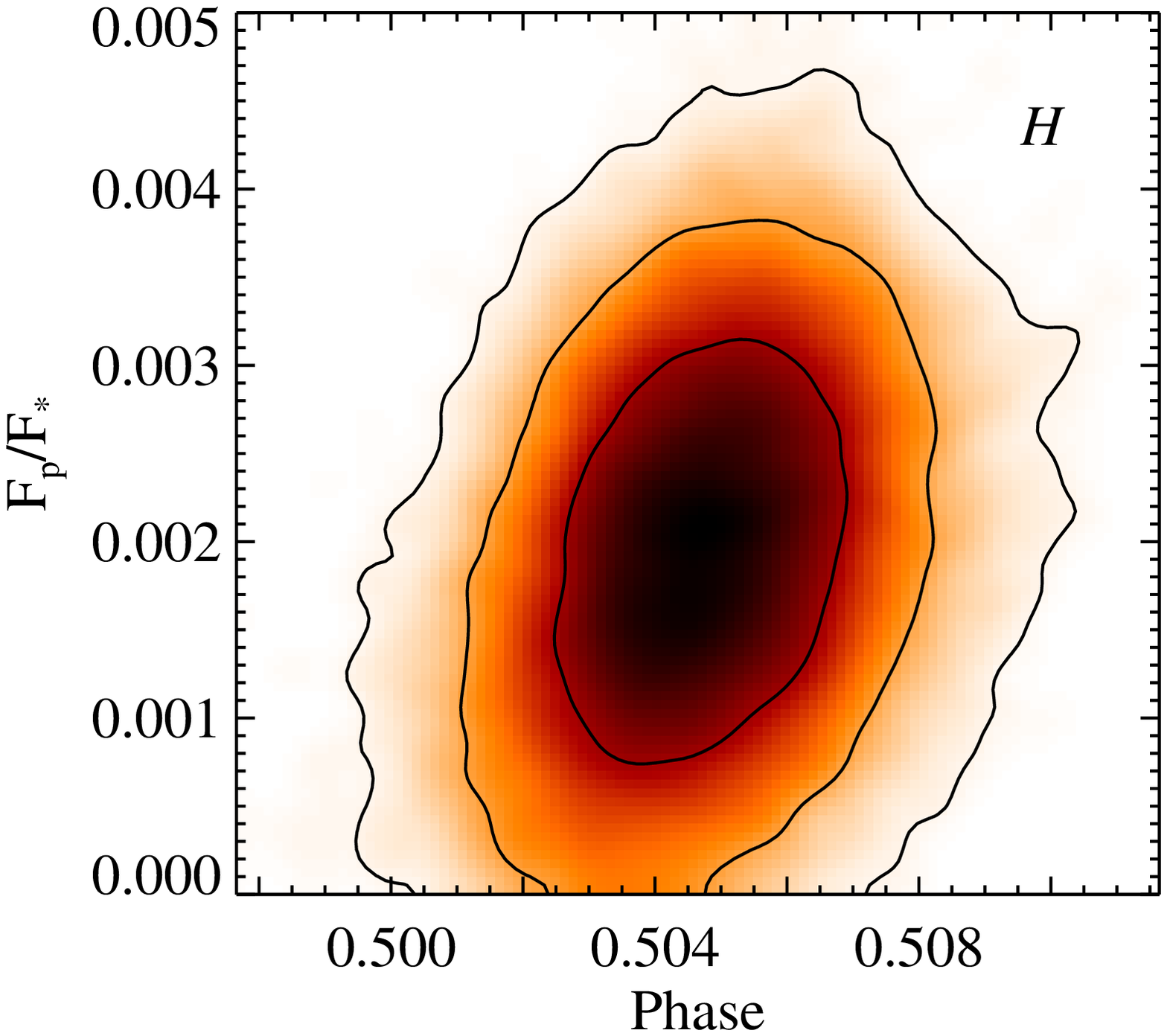}
\includegraphics[width=0.28\hsize]{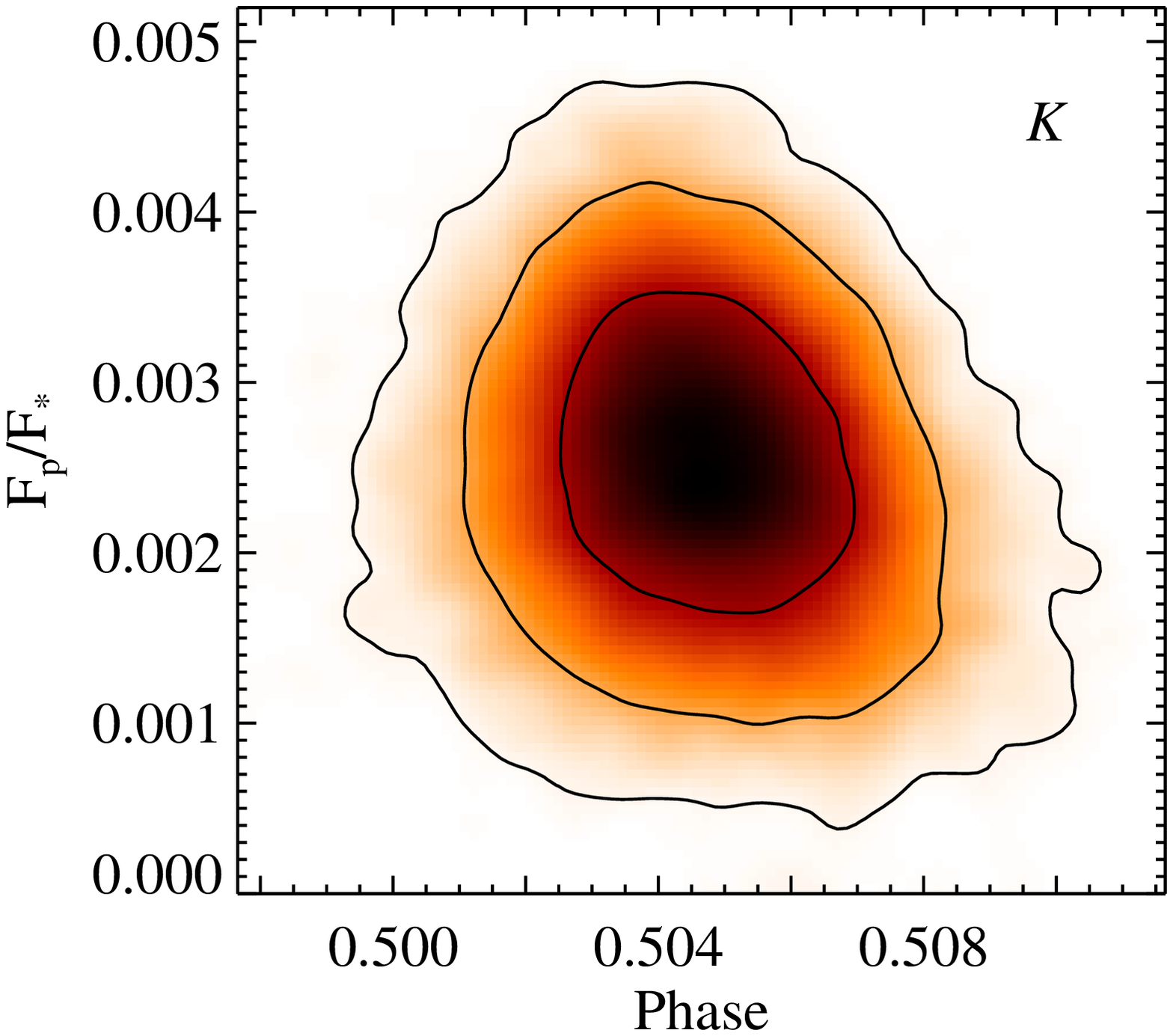}
\caption{Dependence of the measured flux ratio on the measured mid-eclipse time (converted to orbital phase) for the $J$ ({\it left}), $H$ ({\it middle}), and $K$ ({\it right}) bands. Three contour levels correspond to the 1$\sigma$, 2$\sigma$, and 3$\sigma$ confidence levels, respectively. \label{fig:joint_pdf}}
\end{figure*}

The reference-corrected light curves are still systematics dominated. To measure the eclipse depth from the noisy data, we chose to model the light curves using the following combination:
\begin{equation}
f(\mathrm{model})=E(p_i)\left(c_0+\sum\nolimits^{k}_{j=1}c_jv_j\right),
\end{equation}
where $E(p_i)$ is the analytic eclipse model \citep{2002ApJ...580L.171M} for a uniform disk, and the remaining component is the baseline function (BF). In the modeling process, we fit for the mid-eclipse time $T_{\rm{mid,occ}}$ and the planet-to-star flux ratio $F_{\mathrm{p}}/F_{\star}$ in the model $E(p_i)$, along with the baseline coefficients $c_j$ in the BF. The number of $c_j$ depends on the choice of BF model vectors $v_j$, which were empirically determined for each individual light curve. The planetary and stellar parameters needed in the model $E(p_i)$ are taken from \citet{2012MNRAS.422.1988A}, and are listed in Table~\ref{tab:sys}.

We used the Markov chain Monte Carlo (MCMC) technique \citep[e.g.][]{2005AJ....129.1706F,2006ApJ...642..505F} together with the singular-value decomposition (SVD) linear regression method to search for the best-fit values and uncertainties for the free parameters, as described in detail in \citet[][and references therein]{2014A&A...564A...6C,2014A&A...563A..40C}. At each MCMC step, the light-curve data were divided by a perturbed eclipse model, and then the baseline coefficients were linearly solved from the residuals by the SVD method. Since the $g'r'i'z'JHK$ light curves were simultaneously obtained and covered the same secondary eclipse, they were forced to share the same $T_{\rm{mid,occ}}$ in the MCMC modeling process, while they were allowed to have wavelength-dependent $F_{\mathrm{p}}/F_{\star}$. We first ran five chains of 10$^5$ links to derive an initial distribution for each parameter. Then we followed the approach of \citet{2008ApJ...683.1076W} to account for the time-correlated red noise \citep[for details see][]{2014A&A...564A...6C,2014A&A...563A..40C}, which makes use of both the time-averaging method and the prayer-bead method to determine the red-noise scaling factors. The larger one was adopted as $\beta_r$ and was multiplied on the photometric uncertainties. Finally, we ran another five chains of 10$^5$ links on the light curves with rescaled uncertainties to derive the final best-fit values and 68.3\% uncertainties for the free parameters. 

We experimented with a set of baseline function (BF) models to find the one that best characterizes the systematics in each light curve. The best chosen BF models were determined based on the Bayesian information criterion \citep[BIC;][]{Schwarz1978}: $\mathrm{BIC}=\chi^2+k\log(N)$, where $k$ and $N$ refer to the number of free parameters and data points, respectively, and $\chi^2$ is derived based on the photometric uncertainties in which the red-noise effect has not been included. We extensively explored different combinations for the BF vector, including linear, quadratic, and high-order polynomials. Those with the lowest BIC were considered as the final choice. The coefficients of the best BF models are listed in Table~\ref{tab:ap01} and \ref{tab:ap02}. Five of the top-ranked (i.e. with the lowest BIC) BF models are shown in Table~\ref{tab:ap03} for comparison. Different choices of BF model could have an impact on the derived eclipse depth. In general, BF models with high BIC produce eclipse depths that deviate from the chosen one. However, for the BF models with a BIC similar to the lowest BIC, the best-fit values for the eclipse depth, although slightly different, are generally consistent with each other within 1$\sigma$ uncertainties (see Table~\ref{tab:ap03}). 

We also repeated the MCMC modeling process by including the eclipse duration time $T_{58}$ in the light-curve model $E(p_i)$. Given the data quality, we decided to control the values of $T_{\rm{mid,occ}}$ and $F_{\mathrm{p}}/F_{\star}$ with the Gaussian priors. We adopted the derived distributions of $T_{\rm{mid,occ}}$ and $F_{\mathrm{p}}/F_{\star}$ from the aforementioned analysis. The derived eclipse duration time was used to analyze the orbital eccentricity. 

\begin{figure}
\centering
\includegraphics[width=\hsize]{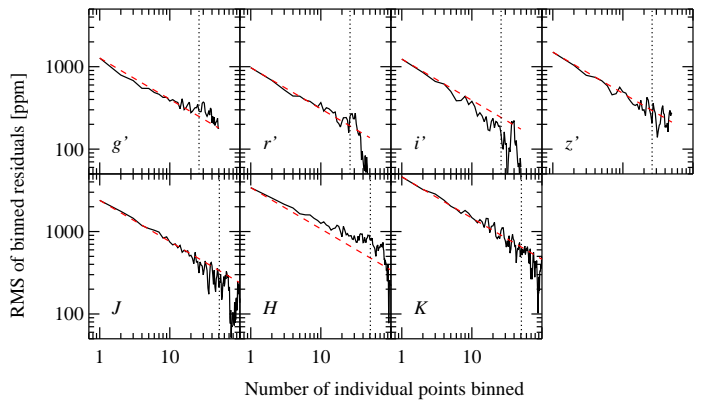}
\caption{Standard deviation of the best-fit light-curve residuals binned in different time resolutions. The vertical dashed line refers to the ingress/egress duration. The red dashed line presents the expected Poisson-like noise. The deviation above the red dashed line indicates time-correlated red noise. \label{fig:rmsbin}}
\end{figure}

We present the derived best-fit parameters in Table~\ref{tab:sys} and \ref{tab:par} along with the adopted parameters and photometric settings. We show the light-curve modeling results in Figs.~\ref{fig:occlc_opt} and \ref{fig:occlc_nir} and the dependence of measured eclipse depth on mid-eclipse time in Fig.~\ref{fig:joint_pdf}. We illustrate the time-averaging process in Fig.~\ref{fig:rmsbin}. After the joint MCMC modeling process, we achieved a photometric precision of 844, 615, 883, 1035, 1117, 1764, and 2325~ppm in two-minute intervals in the best-fit light-curve residuals for the $g'r'i'z'JHK$ bands, respectively. The corresponding photon noise limits in two-minute intervals are $3.2\times10^{-4}$, $2.4\times10^{-4}$, $3.3\times10^{-4}$, $3.3\times10^{-4}$, $3.4\times10^{-4}$, $3.4\times10^{-4}$, and $5.6\times10^{-4}$, respectively.  

   \begin{table*}
     \small
     \centering
     \caption{Flux ratios and other parameters from the joint analysis}
     \label{tab:par}
     \begin{tabular}{cccccccc}
       \hline\hline\noalign{\smallskip}
       Filter & $F_{\mathrm{p}}/F_{\star}$ & 3$\sigma$ upper limit & \multicolumn{2}{c}{$A_g$} & $T_B$ & $N_{\rm{ref}}$\tablefootmark{a} & Aperture\tablefootmark{b}\\\noalign{\smallskip}\cline{4-5}\noalign{\smallskip}
              &  (\%) &  (\%) & effective & corrected & (K) & & (pix)\\\noalign{\smallskip}
       \hline\noalign{\smallskip}
       $g'$ & 0.049 $^{+0.030}_{-0.027}$ & $<$0.138 & 0.75 $^{+0.46}_{-0.41}$ 
            & 0.73 $^{+0.46}_{-0.41}$ & 3326 $^{+186}_{-273}$ & 8 & $(22,36,42)$\\\noalign{\smallskip}
       $r'$ & 0.040 $^{+0.016}_{-0.016}$ & $<$0.089 & 0.62 $^{+0.24}_{-0.25}$ 
            & 0.48 $^{+0.24}_{-0.25}$ & 2840 $^{+125}_{-172}$ & 9 & $(25,33,40)$\\\noalign{\smallskip}
       $i'$ & 0.047 $^{+0.034}_{-0.029}$ & $<$0.147 & 0.71 $^{+0.51}_{-0.44}$ 
            & 0.35 $^{+0.51}_{-0.44}$ & 2614 $^{+214}_{-309}$ & 8 & $(20,36,42)$\\\noalign{\smallskip}
       $z'$ & 0.017 $^{+0.020}_{-0.012}$ & $<$0.096 & 0.25 $^{+0.31}_{-0.19}$ 
            & $<$0.77 & 2079 $^{+235}_{-287}$ & 9 & $(18,36,42)$\\\noalign{\smallskip}
       $J$  & 0.129 $^{+0.055}_{-0.055}$ & $<$0.295 & 1.97 $^{+0.84}_{-0.84}$ 
            & ... & 2453 $^{+198}_{-258}$ & 3 & $(6.5,13.5,22)$\\\noalign{\smallskip}
       $H$  & 0.194 $^{+0.078}_{-0.078}$ & $<$0.431 & 2.96 $^{+1.20}_{-1.19}$ 
            & ... & 2462 $^{+245}_{-302}$ & 3 & $(6.5,15.5,24)$\\\noalign{\smallskip}
       $K$  & 0.253 $^{+0.063}_{-0.060}$ & $<$0.442 & 3.86 $^{+0.96}_{-0.92}$ 
            & ... & 2306 $^{+177}_{-187}$ & 3 & $(3.5,3.5,18)$\\\noalign{\smallskip}
       \hline
     \end{tabular}\\
     \tablefoot{
        \tablefoottext{a}{Number of reference stars that are used to create the composite reference. 
                          Note that when different bands have the same $N_{\rm{ref}}$, they do not 
                          necessarily have the same reference star ensemble.  
                          }
        \tablefoottext{b}{Bracketed numbers refer to aperture radius, inner sky annulus radius, and 
                          outer sky annulus radius, respectively.} 
                }
   \end{table*}

\section{Results and discussion}\label{sec:discuss}

We detect the secondary-eclipse dip in the $K$ band with a flux ratio of 0.253 $^{+0.063}_{-0.060}$\% at the 4.2$\sigma$ significance level, and tentatively detect the dip in the $J$ (2.5$\sigma$) and $H$ (2.3$\sigma$) bands with flux ratios of 0.194 $\pm$ 0.078\% and 0.129 $\pm$ 0.055\%, respectively.  In the optical bands, we do not detect the dip in the $z'$ band, while we tentatively detect the dip in the $g'r'i'$ bands at very low significance level.

\subsection{Orbital eccentricity}\label{sec:ecc}

The orbital eccentricity $e$ and the argument of periastron $\omega$ are linked by the central times and the durations of both transit and eclipse, see for example, Eq.~(17) and (18) in \citet{2009ApJ...698.1778R}. The center of the secondary eclipse is expected to occur at an observed phase $\phi$ = 0.5002, if a circular orbit for \object{WASP-46b} is assumed. This calculation has accounted for the light travel time of 24.3~s in the system \citep{2005ApJ...623L..45L}. From our joint analysis of the seven eclipse light curves, we derive a mid-eclipse time that deviates 9.6 $^{+3.0}_{-2.9}$~minutes from this expected time, which corresponds to a corrected phase of 0.5047 $\pm$ 0.0014. This offset time corresponds to $e\cos\omega=0.0073\pm0.0022$. In addition, we measure an eclipse duration of 0.0639 $^{+0.0049}_{-0.0044}$\,days that is marginally shorter than the transit duration ($T_{14}=0.06973$\,days), corresponding to $e\sin\omega=-0.018\pm0.015$. This calculation gives an eccentricity of $e=0.020^{+0.014}_{-0.010}$ at low significance, which is consistent with the value derived from the radial velocity data \citep[$e=0.018^{+0.021}_{-0.013}$;][]{2012MNRAS.422.1988A}. The eccentricity is consistent with zero at 2$\sigma$ level. It is possible that the orbit is indeed slightly eccentric. However, we also caution that our central eclipse time and eclipse duration could be contaminated by the time-correlated systematics. To better constrain these orbital parameters, light curves with higher precision and deeper dips (i.e. higher signal-to-noise ratio) are required, for instance, from warm {\it Spitzer}. 

\subsection{Near-infrared thermal emission from the dayside atmosphere of WASP-46b}\label{sec:nirspec}

\begin{figure}
\centering
\includegraphics[width=0.98\hsize]{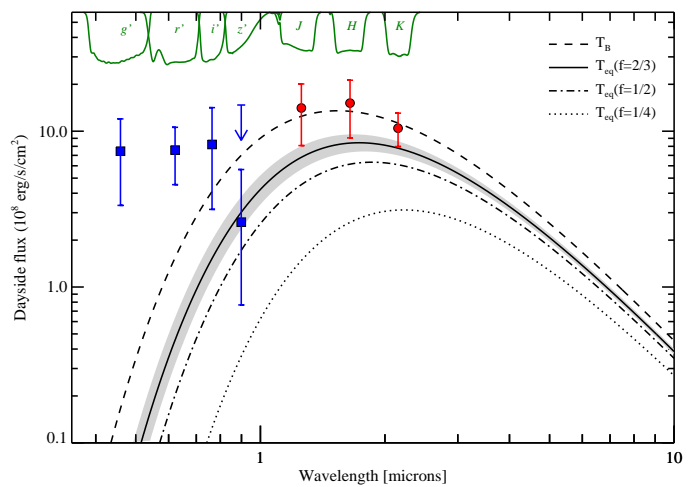}
\caption{Dayside flux measurements of \object{WASP-46b} compared with blackbody spectral energy distributions. Our measurements are shown as blue squares for the optical bands and as red circles for the near-infrared bands. The blue arrow shows the 3$\sigma$ upper limit in the $z'$ band. Blackbody profiles of four temperatures are shown, including the $JHK$-weighted brightness temperature $T_B$ and three typical equilibrium temperatures (i.e. with $f=1/4,1/2,2/3$, corresponding to planet-wide, dayside, no heat redistribution, respectively). The shaded area shows the uncertainties of the equilibrium temperature.\label{fig:sed}}
\end{figure}

The derived $J$-, $H$-, $K$-band flux ratios of 0.129 $\pm$ 0.055\%, 0.194 $\pm$ 0.078\%, 0.253 $^{+0.063}_{-0.060}$\% can be translated into brightness temperatures of 2453 $^{+198}_{-258}$~K, 2462 $^{+245}_{-302}$~K, 2306 $^{+177}_{-187}$~K, respectively. In this calculation, we simply assumed that the planet radiates blackbody emission. We interpolated the stellar spectrum based on the Kurucz models \citep{1979ApJS...40....1K}, using the stellar parameters $T_{\rm{eff,\star}}=5620$\,K, $\rm{[Fe/H]}=-0.37$, and $\log g_{\star}=4.493$ \citep{2012MNRAS.422.1988A}. After individually integrating the planetary emission and the stellar emission over the bandpass of each filter and taking the planet-to-star area ratio ($R_{\rm{p}}^2/R_{\star}^2=0.02155$) into account, we record the brightness temperature that results in the flux ratio that best matches the measured value. The derived brightness temperatures are significantly higher ($\sim$3$\sigma$) than the equilibrium temperature of 1654 $\pm$ 50~K that assumes homogeneous planet-wide redistribution of the incident heat ($A_B=0$ and $f=1/4$). The equilibrium temperature is given by the equation
\begin{equation}
T_{\mathrm{eq}}=T_{\mathrm{eff,\star}}\sqrt{R_{\star}/a}[f(1-A_B)]^{1/4},
\end{equation}
where $A_B$ is the Bond albedo, and $f$ ranges from 1/4 to 2/3 \citep{2011ApJ...729...54C}. Assuming $A_B=0$ and $f=2/3$, that is, instant re-emission of incident heat with no redistribution, the highest equilibrium temperature is calculated to be 2120 $\pm$ 67~K, which agrees with the near-infrared brightness temperatures at 1$\sigma$ level. Therefore, if the energy excess is not of physical origin (e.g. internal energy sources), it could be explained by the large uncertainties stemming from the poor data quality. This also indicates that the energy redistribution efficiency in the dayside atmosphere is very poor. 

In Fig.~\ref{fig:sed}, we construct a spectral energy distribution for the thermal emission of \object{WASP-46b} based on our eclipse-depth measurements, and we show the corresponding flux-ratio spectrum in Fig.~\ref{fig:occspec}. The near-infrared measurements can be explained by an isothermal temperature profile of 2386~K, which is the weighted mean of the three brightness temperatures. The $J$, $H$, and $K$ bands probe deep into the lower layers of the atmosphere because of weaker molecular absorptions at these bands. The isothermal temperature profile is expected to be present at the optical depth $\tau\sim1$ \citep{2008ApJS..179..484H,2009ApJ...707...24M,2010A&A...520A..27G}. 

Given the high temperature down in the atmosphere, if there is no internal energy source, it is unlikely that a strong thermal inversion exists in the upper atmosphere, otherwise the energy balance would be violated. The sign of temperature decreasing towards higher layer can tentatively be seen in the $J$, $H$, and $K$ bands. Although temperatures in the three bands agree well with each other within the 1$\sigma$ error bar and the best-fit temperatures of the $J$ and $H$ bands are almost the same, the best-fit $K$-band temperature is $\sim$150~K lower. According to the near-infrared opacity curve (dominated by water), the $J$, $H$, and $K$ bands probe the pressure layers that gradually increase \citep[e.g.][]{2008ApJ...676L..61B,2008ApJ...678.1419F}. This indicates that the higher layer that is probed by the $K$ band has a slightly lower temperature than the lower layer that is probed by the $J$ and $H$ bands. 

On the other hand, the high equilibrium temperature (1654--2120\,K, assuming $A_B=0$ and $1/4\leq f\leq 2/3$) possibly allows gaseous TiO and VO to be present in the upper atmosphere because Ti and V start to condense at 1670~K at 1~mbar \citep{2008ApJ...678.1419F}. However, the potential upper-atmosphere absorbers, such as the TiO and VO compounds or the sulfur compounds \citep{2009ApJ...701L..20Z} or some unknown stratospheric absorbers, might have been destroyed by the high UV flux of the host star if the hypothesis proposed by \citet{2010ApJ...720.1569K} applies. The discovery paper \citep{2012MNRAS.422.1988A} has found a 16-day rotational modulation and weak \ion{Ca}{II} H+K for the host star. A recent study led by \citet{2013ApJ...766....9S} furthermore derived a high far-UV fractional luminosity $\log(L_{FUV}/L_{bol})=-4.546\pm0.079$ based on the NASA Galaxy Evolution Explorer (GALEX) measurements. The possible absence of high-altitude absorbers would indicate that there is no strong thermal inversion. 

As a next step, more repeated near-infrared observations are needed to improve the confidence level of the measured eclipse depth, allowing us to accurately derive the heat redistribution efficiency. Furthermore, future measurements at longer wavelength, for example, warm {\it Spitzer} 3.6~$\mu$m and 4.5~$\mu$m, are crucial to construct the pressure-temperature profile and to decompose the molecular contents. 

\subsection{Reflected light from the dayside atmosphere of WASP-46b?}\label{sec:optspec}

\begin{figure}
\centering
\includegraphics[width=\hsize]{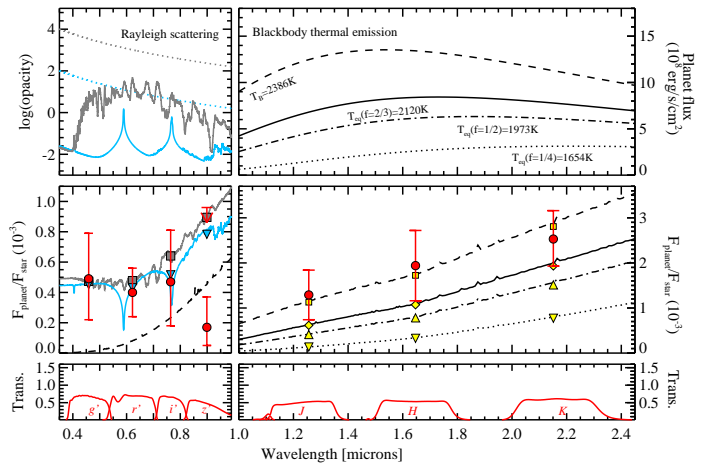}
\caption{Dayside planet-to-star flux ratios of \object{WASP-46b} compared with theoretical models. The optical and NIR are divided into {\it left} and {\it right} panels for better display scale. The {\it top panels} show the assumed theoretical models. Rayleigh scattering and atomic/molecular absorption profiles are assumed for the reflected light, where a Na/K model (blue) and a TiO/VO model (gray) from \citet{2008ApJ...678.1419F,2010ApJ...709.1396F} are adopted for the absorption profiles. Blackbody profiles are assumed for the thermal emission light. The {\it middle panels} show the corresponding planet-to-star flux ratios. Our data points are shown in red circles with error bars, while the values that are integrated in each bandpass are shown in yellow (blue or gray for the optical) shapes. The {\it bottom panels} show the transmission curve of each band. \label{fig:occspec}}
\end{figure}

Presuming that the tentative eclipse detections in the optical bands are real planetary signals instead of systematic false positives, we investigated what atmosphere they could deliver. As shown in Fig.~\ref{fig:sed}, the optical measurements start to deviate from the blackbody emission profile blueward of the $z'$ band. The non-detection of the $z'$-band eclipse dip indicates that the emission at this band could be very low, while its 3$\sigma$ upper limit in principal allows the 2386~K isothermal temperature profile. In contrast, the best-fit values of the $g'$-, $r'$-, $i'$-band measurements are clearly off the blackbody profile.

We derived the effective geometric albedos as listed in Table~\ref{tab:par}. The geometric albedo is defined as $A_g=(F_{\mathrm{p}}/F_{\star})/(R_{\mathrm{p}}/a)^2$. Assuming a blackbody emission of 2386~K, we corrected the thermal emission portion in the optical geometric albedos and list them in Table~\ref{tab:par} as well. The values of these corrected geometric albedos appear to be very high. However, their large uncertainties degrade the significance, thus allowing the actual geometric albedos to be possibly lower. The quality of our measurements does not seem to warrant a detailed modeling of the optical spectrum. In Fig.~\ref{fig:occspec}, disregarding our near-infrared measurements, we construct two optical toy spectra composed of both reflection and emission for illustration purpose. The reflection portion was generated following the approach described by \citet{2013ApJ...772L..16E}, in which a Rayleigh scattering profile and two absorption profiles of the \citet{2008ApJ...678.1419F,2010ApJ...709.1396F} models were combined to simulate the reflected light. Under the current simple assumption, the reflected light could account for the high flux ratios in the optical bands, indicating a high-altitude reflective cloud. However, since our near-infrared high brightness temperatures already suggest a very low Bond albedo, the high geometric albedos derived here are probably not real, unless the actual near-infrared eclipse depths are far shallower than the values measured here. Future transit spectroscopic observations in the optical as well as higher-precision near-infrared secondary-eclipse observations will be very useful to resolve this problem.

\section{Conclusions}\label{sec:con}

We observed one secondary eclipse of \object{WASP-46b} simultaneously in the $g'r'i'z'JHK$ bands. We detected or tentatively detected the thermal emission in the $J$, $H$, $K$ bands with a flux ratio of 0.129 $\pm$ 0.055\%, 0.194 $\pm$ 0.078\%, 0.253 $^{+0.063}_{-0.060}$\%, respectively. Corresponding brightness temperatures of 2453 $^{+198}_{-258}$~K, 2462 $^{+245}_{-302}$~K, 2306 $^{+177}_{-187}$~K are consistent with an isothermal temperature profile of 2386~K. The slightly lower $K$-band temperature might indicate a decreasing temperature profile towards higher atmospheric layer. While the best-fit near-infrared brightness temperatures are all higher than the expected highest equilibrium temperature assuming zero albedo and no heat redistribution, they are still nearly consistent with the expected highest temperature at 1$\sigma$ level, given their large uncertainties. Furthermore, we derived eclipse depths for the four optical bands. After correcting for the thermal emission portion, the geometric albedos still appear to be high, which is inconsistent with the low Bond albedo inferred from the near-infrared measurements. Future higher-precision observations are required to confirm our detections. 

\begin{acknowledgements}
We are grateful to the anonymous referee for the careful reading and helpful comments that improved the manuscript. We thank Luigi Mancini for providing advice to improve this manuscript. G.C. acknowledges Chinese Academy of Sciences and Max Planck Society for the support of the doctoral training program. H.W. acknowledges the support by NSFC grants 11173060, 11127903, and 11233007. This work is supported by the Strategic Priority Research Program "The Emergence of Cosmological Structures" of the Chinese Academy of Sciences, Grant No. XDB09000000. Part of the funding for GROND (both hardware 84 and personnel) was generously granted by the Leibniz-Prize to G. Hasinger 85 (DFG grant HA 1850/28-1).
\end{acknowledgements}

\bibliographystyle{aa} 
\bibliography{ref} 

\clearpage

  \Online

  \begin{appendix}
  \section{Light-curve baseline model selection}\label{sec:ap01}
  Tables~\ref{tab:ap01} and \ref{tab:ap02} list the baseline coefficients of the chosen models for the seven light curves. Table~\ref{tab:ap03} gives five of the top-ranked baseline function models based on their BICs. The models that are numbered with one are the best chosen ones.

   \begin{table}
     \scriptsize
     \centering
     \caption{Baseline coefficients of the chosen models for the $g'r'i'z'$ light curves}
     \label{tab:ap01}
     \begin{tabular}{crrrr}
     \hline\hline\noalign{\smallskip}
     Coeff. & \multicolumn{1}{c}{$g'$} & \multicolumn{1}{c}{$r'$} & \multicolumn{1}{c}{$i'$} & \multicolumn{1}{c}{$z'$}\\\noalign{\smallskip}
     \hline\noalign{\smallskip}
       $c_0$ & 0.99997 $^{(12)}_{(11)}$  & 1.000066 $^{(72)}_{(72)}$ 
             & 0.99941 $^{(27)}_{(23)}$  & 1.000049 $^{(93)}_{(56)}$ \\\noalign{\smallskip}
       $c_1$ & $-$0.0000240 $^{(36)}_{(32)}$ & $-$0.01157 $^{(37)}_{(46)}$ 
             & 0.0144 $^{(21)}_{(17)}$   & $-$0.00284 $^{(24)}_{(42)}$ \\\noalign{\smallskip}
       $c_2$ & $-$0.01731 $^{(36)}_{(24)}$ & \multicolumn{1}{c}{...} 
             & 0.107 $^{(09)}_{(10)}$    & \multicolumn{1}{c}{...}\\\noalign{\smallskip}
     \hline
     \end{tabular}
     \tablefoot{
                The two bracketed digits in super-/subscript show the 68.3\% 
                uncertainties of corresponding best-fit values. They should be 
                compared to the last two digits of the best-fit values.
                }
   \end{table}

   \begin{table}
     \small
     \centering
     \caption{Baseline coefficients of the chosen models for the $JHK$ light curves}
     \label{tab:ap02}
     \begin{tabular}{crrr}
     \hline\hline\noalign{\smallskip}
     Coeff. & \multicolumn{1}{c}{$J$} & \multicolumn{1}{c}{$H$} & \multicolumn{1}{c}{$K$}\\\noalign{\smallskip}
     \hline\noalign{\smallskip}
       $c_0$ & 0.99706 $^{(50)}_{(50)}$   & 0.99811 $^{(57)}_{(56)}$ 
             & 0.99761 $^{(27)}_{(26)}$\\\noalign{\smallskip}
       $c_1$ & 0.002053 $^{(15)}_{(15)}$  & $-$0.0039502 $^{(56)}_{(32)}$ 
             & $-$0.00747 $^{(43)}_{(40)}$\\\noalign{\smallskip}
       $c_2$ & $-$0.003207 $^{(17)}_{(26)}$ & $-$0.00126 $^{(27)}_{(23)}$ 
             & \multicolumn{1}{c}{...}\\\noalign{\smallskip}
       $c_3$ & 0.003360 $^{(52)}_{(52)}$  & 0.00272 $^{(16)}_{(12)}$ 
             & \multicolumn{1}{c}{...}\\\noalign{\smallskip}
       $c_4$ & 0.0631 $^{(28)}_{(26)}$    & 0.00382 $^{(55)}_{(56)}$ 
             & \multicolumn{1}{c}{...}\\\noalign{\smallskip}
       $c_5$ & 1.34 $^{(24)}_{(25)}$      & 0.00758 $^{(34)}_{(33)}$ 
             & \multicolumn{1}{c}{...}\\\noalign{\smallskip}
       $c_6$ & $-$33.32 $^{(2.4)}_{(1.8)}$  & 0.0715 $^{(17)}_{(19)}$ 
             & \multicolumn{1}{c}{...}\\\noalign{\smallskip}
     \hline
     \end{tabular}
     \tablefoot{
                The two bracketed digits in super-/subscript show the 68.3\% 
                uncertainties of corresponding best-fit values. They should be 
                compared to the last two digits of the best-fit values.
                }
   \end{table}

   \begin{table}
     \scriptsize
     \centering
     \caption{Top-ranked light-curve baseline models}
     \label{tab:ap03}
     \begin{tabular}{cccccl}
     \hline\hline\noalign{\smallskip}
     Model & $\Delta$BIC & RMS & $\beta_r$ & $F_{\rm{p}}/F_{\star}$\tablefootmark{a} & Model vectors $v_j$\\\noalign{\smallskip}
     (\#) & & (ppm) & & (\%) &\\\noalign{\smallskip}
     \hline\noalign{\smallskip}
       \multicolumn{6}{c}{\smallskip$g'$ band\dotfill}\\
       1 & 0   & 1271 & 1.29 & 0.050 $^{+0.022}_{-0.021}$ & $\{s_y,t\}$\\\noalign{\smallskip}
       2 & 1.3 & 1275 & 1.23 & 0.052 $^{+0.021}_{-0.022}$ & $\{s_x,t\}$\\\noalign{\smallskip}
       3 & 3.7 & 1266 & 1.32 & 0.044 $^{+0.022}_{-0.021}$ & $\{x,s_y,t\}$\\\noalign{\smallskip}
       4 & 4.0 & 1267 & 1.31 & 0.045 $^{+0.022}_{-0.021}$ & $\{x,s,t\}$\\\noalign{\smallskip}
       5 & 4.8 & 1270 & 1.29 & 0.049 $^{+0.022}_{-0.021}$ & $\{y,s_y,t\}$\\\noalign{\smallskip}
       \multicolumn{6}{c}{$r'$ band\dotfill}\\\noalign{\smallskip}
       1 & 0   & 974 & 1.01 & 0.042 $^{+0.015}_{-0.015}$ & $\{t\}$\\\noalign{\smallskip}
       2 & 0.8 & 976 & 1.02 & 0.034 $^{+0.015}_{-0.015}$ & $\{x\}$\\\noalign{\smallskip}
       3 & 3.7 & 969 & 1.15 & 0.033 $^{+0.017}_{-0.016}$ & $\{s_x,t\}$\\\noalign{\smallskip}
       4 & 4.6 & 972 & 1.01 & 0.039 $^{+0.016}_{-0.016}$ & $\{x,t\}$\\\noalign{\smallskip}
       5 & 5.0 & 973 & 1.06 & 0.039 $^{+0.018}_{-0.017}$ & $\{s_y,t\}$\\\noalign{\smallskip}
       \multicolumn{6}{c}{$i'$ band\dotfill}\\\noalign{\smallskip}
       1 & 0   & 1237 & 1.00 & 0.050 $^{+0.033}_{-0.029}$ & $\{t,t^2\}$\\\noalign{\smallskip}
       2 & 0.2 & 1237 & 1.00 & 0.053 $^{+0.032}_{-0.029}$ & $\{t,z\}$\\\noalign{\smallskip}
       3 & 3.4 & 1231 & 1.00 & 0.051 $^{+0.033}_{-0.029}$ & $\{s_x,t,t^2\}$\\\noalign{\smallskip}
       4 & 3.4 & 1231 & 1.00 & 0.058 $^{+0.032}_{-0.031}$ & $\{s_y,t,z\}$\\\noalign{\smallskip}
       5 & 3.5 & 1231 & 1.00 & 0.057 $^{+0.034}_{-0.031}$ & $\{s_y,t,t^2\}$\\\noalign{\smallskip}
       \multicolumn{6}{c}{$z'$ band\dotfill}\\\noalign{\smallskip}
       1 & 0   & 1499 & 1.33 & 0.012 $^{+0.015}_{-0.008}$ & $\{z\}$\\\noalign{\smallskip}
       2 & 0.8 & 1502 & 1.69 & 0.010 $^{+0.013}_{-0.007}$ & $\{t\}$\\\noalign{\smallskip}
       3 & 1.4 & 1504 & 1.37 & 0.012 $^{+0.015}_{-0.009}$ & $\{s_x\}$\\\noalign{\smallskip}
       4 & 2.2 & 1508 & 1.95 & 0.009 $^{+0.012}_{-0.007}$ & $\{x\}$\\\noalign{\smallskip}
       5 & 2.4 & 1508 & 1.55 & 0.011 $^{+0.015}_{-0.008}$ & $\{s_y\}$\\\noalign{\smallskip}
       \multicolumn{6}{c}{$J$ band\dotfill}\\\noalign{\smallskip}
       1 & 0   & 2390 & 1.00 & 0.136 $^{+0.055}_{-0.054}$ & $\{x,y,s_y,t,t^2,t^3\}$\\\noalign{\smallskip}
       2 & 2.4 & 2356 & 1.00 & 0.139 $^{+0.055}_{-0.055}$ & $\{x,y,xy,y^2,s_y,t,t^2,t^3\}$\\\noalign{\smallskip}
       3 & 4.2 & 2384 & 1.00 & 0.139 $^{+0.056}_{-0.055}$ & $\{x,y,xy,s_y,t,t^2,t^3\}$\\\noalign{\smallskip}
       4 & 5.6 & 2389 & 1.00 & 0.131 $^{+0.057}_{-0.055}$ & $\{x,y,s_x,s_y,t,t^2,t^3\}$\\\noalign{\smallskip}
       5 & 8.2 & 2356 & 1.00 & 0.136 $^{+0.057}_{-0.055}$ & $\{x,y,xy,y^2,s_x,s_y,t,t^2,t^3\}$\\\noalign{\smallskip}
       \multicolumn{6}{c}{$H$ band\dotfill}\\\noalign{\smallskip}
       1 & 0   & 3407 & 1.43 & 0.199 $^{+0.056}_{-0.057}$ & $\{x,y,xy,y^2,s,z\}$\\\noalign{\smallskip}
       2 & 1.1 & 3414 & 1.45 & 0.184 $^{+0.056}_{-0.056}$ & $\{x,y,xy,y^2,s_y,z\}$\\\noalign{\smallskip}
       3 & 3.5 & 3395 & 1.44 & 0.205 $^{+0.058}_{-0.057}$ & $\{x,y,xy,x^2,y^2,s,z\}$\\\noalign{\smallskip}
       4 & 3.8 & 3456 & 1.55 & 0.185 $^{+0.068}_{-0.068}$ & $\{x,y,s,t,z\}$\\\noalign{\smallskip}
       5 & 5.1 & 3490 & 1.62 & 0.194 $^{+0.069}_{-0.067}$ & $\{x,t,s,z\}$\\\noalign{\smallskip}
       \multicolumn{6}{c}{$K$ band\dotfill}\\\noalign{\smallskip}
       1 & 0   & 4612 & 1.00 & 0.262 $^{+0.060}_{-0.060}$ & $\{s_y\}$\\\noalign{\smallskip}
       2 & 4.4 & 4603 & 1.00 & 0.276 $^{+0.061}_{-0.061}$ & $\{y,s_y\}$\\\noalign{\smallskip}
       3 & 5.3 & 4608 & 1.00 & 0.269 $^{+0.060}_{-0.061}$ & $\{s_y,t\}$\\\noalign{\smallskip}
       4 & 5.7 & 4610 & 1.00 & 0.260 $^{+0.060}_{-0.061}$ & $\{s_y,z\}$\\\noalign{\smallskip}
       5 & 5.9 & 4612 & 1.00 & 0.263 $^{+0.061}_{-0.061}$ & $\{x,s_y\}$\\\noalign{\smallskip}
     \hline
     \end{tabular}
     \tablefoot{
                \tablefoottext{a}{The listed 1$\sigma$ uncertainties were derived 
                                  before the inclusion of red noise.}
                }
   \end{table}

  \end{appendix}

\end{document}